\documentclass[aps,pra,reprint,amsmath,amssymb,superscriptaddress,showkeys]{revtex4-1}

\usepackage[english]{babel}
\usepackage{graphicx}% Include figure files
\usepackage{bm}% bold math
\usepackage{url}
\usepackage{siunitx}
\usepackage{hyperref}
\usepackage[hyphenbreaks]{breakurl}

\begin{document}

\preprint{APS/123-QED}

\title{An Exercise in Open Data: Triple Axis Data on Si single crystal}
\thanks{Authors in alphabetical order}%

\author{Pavlo Baloh}
\affiliation{Pavol Jozef \v{S}af\'arik University, Ko\v{s}ice, Slovakia}
%\email{pavlo.baloh@student.upjs.sk}

\author{Lukas Bauer}
\affiliation{Forschungszentrum J\"ulich GmbH, J\"ulich, D-85748 Garching, Germany}
%\email{l.bauer@fz-juelich.de}

\author{Lukas Beddrich}
\affiliation{Heinz Maier-Leibnitz Zentrum (MLZ), Technische Universit\"at M\"unchen, D-85748 Garching, Germany}
%\email{Lukas.Beddrich@frm2.tum.de}

\author{Ane\v{z}ka Bendov\'a}
\affiliation{Charles University, Faculty of Mathematics and Physics, Department of Condensed Matter Physics, Prague, Czech republic}
%\email{anezka.bendova@gmail.com}

\author{Petr \v{C}erm\'ak}
\affiliation{Charles University, Faculty of Mathematics and Physics, Department of Condensed Matter Physics, Prague, Czech republic}
%\email{cermak@mag.mff.cuni.cz}

\author{Korbinian Fellner}
\affiliation{Heinz Maier-Leibnitz Zentrum (MLZ), Technische Universit\"at M\"unchen, D-85748 Garching, Germany}
%\email{korbinian@fellner-home.de}

\author{Madhu Ghanathe}
\affiliation{Heinz Maier-Leibnitz Zentrum (MLZ), Technische Universit\"at M\"unchen, D-85748 Garching, Germany}
%\email{madhubghanathe@gmail.com}

\author{Octavio Emmanuel Hern\'andez Alvarez}
\affiliation{Charles University, Faculty of Mathematics and Physics, Department of Condensed Matter Physics, Prague, Czech republic}
%\email{octavio.hernandezaz@udlap.mx}

\author{\v{S}tefan Hricov}
\affiliation{Charles University, Faculty of Science, Prague, Czech republic}
%\email{hricovs@natur.cuni.cz}

\author{Johanna K. Jochum}
\affiliation{Heinz Maier-Leibnitz Zentrum (MLZ), Technische Universit\"at M\"unchen, D-85748 Garching, Germany}
%\email{jjochum@frm2.tum.de}

\author{Liliia Kotvytska}
\affiliation{Pavol Jozef \v{S}af\'arik University, Ko\v{s}ice, Slovakia}
%\email{ramszovk@natur.cuni.cz}

\author{Sonu Kumar}
\affiliation{Charles University, Faculty of Mathematics and Physics, Department of Condensed Matter Physics, Prague, Czech republic}
\affiliation{Adam Mickiewicz University, Faculty of Physics, Poznan, Poland}
%\email{sonuwhf@gmail.com}

\author{Ankit Labh}
\affiliation{Charles University, Faculty of Mathematics and Physics, Department of Condensed Matter Physics, Prague, Czech republic}
%\email{ankitlabh@gmail.com}

\author{Petr Machovec}
\affiliation{Charles University, Faculty of Mathematics and Physics, Department of Condensed Matter Physics, Prague, Czech republic}
%\email{machovec9@seznam.cz}

\author{Brian R. Pauw}
\affiliation{Bundesanstalt f\"ur Materialforschung und Pr\"ufung, Berlin, Germany}
%\email{Brian.Pauw@bam.de}

\author{Klaudie Ramszov\'a}
\affiliation{Charles University, Faculty of Science, Prague, Czech republic}
%\email{ramszovk@natur.cuni.cz}

\author{Erik Walz}
\affiliation{Heinz Maier-Leibnitz Zentrum (MLZ), Technische Universit\"at M\"unchen, D-85748 Garching, Germany}
%\email{erik.walz@frm2.tum.de}

\author{Peter Wild}
\affiliation{Heinz Maier-Leibnitz Zentrum (MLZ), Technische Universit\"at M\"unchen, D-85748 Garching, Germany}
%\email{peter.wild@frm2.tum.de}

\collaboration{Czech-Bavarian Mini-School on large scale facilities and open data 2022}%\noaffiliation
\noaffiliation

\date{\today}% It is always \today, today,

\begin{abstract}

Efforts are rising in opening up science by making data more transparent and more easily available, including the data reduction and evaluation procedures and code. 
A strong foundation for this is the F.A.I.R. principle, building on Findability, Accessibility, Interoperability, and Reuse of digital assets, complemented by the letter T for trustworthyness of the data. 
Here, we have used data, which was made available by the Institute Laue-Langevin and can be identified using a DOI, to follow the F.A.I.R.+T. principle in extracting, evaluating and publishing triple axis data, recorded at IN3.

\end{abstract}

%Use showkeys class option if keyword
\maketitle

\section{Introduction}

Disclaimer: This document is the second iteration of the manuscript published in 2020 on arXiv.org\cite{minischool2020}.
A new group of students continued the project started in 2020 during the first ``Czech-Bavarian mini-school on large-scale facilites and open data'' and added their findings to this iteration of the manuscript.\\

Open Science is defined as \emph{``the practice of science in such a way that others can collaborate and contribute, where research data, lab notes and other research processes are freely available, under terms that enable reuse, redistribution and reproduction of the research and its underlying data and methods''} \cite{foster}.\\

There is a rising demand around the world for open science and many organisations are putting efforts into increasing the amount of infrastructures available for open data \cite{Bilbao, NIST, plos, sharing, panosc, nfdi, expands}. 
It is especially the scientists working at large-scale or other user support facilities that are pushing forward this endeavour in the natural sciences \cite{panosc, nfdi, expands}.
This is understandable, considering these facilities do produce unique, and therefore very valuable data sets, which cannot easily be reproduced due to limited measurement time. As a result, these data sets should be made available (after an embargo time) to the entire scientific community. 

Most large-scale facilities offer measurement time in a proposal based system, where proposals are rated and ranked and only very few proposals are granted measurement time.
It is therefore highly unlikely that the same experiment will be performed for a second time. 
Unfortunately, it happens rather often that during a PhD project, data is measured at large-scale facilities, that will not be evaluated or published anymore by the respective PhD student, or another member of the research group.
In such cases, data that has been taken at large-scale facilities is stored away somewhere, unpublished.
If such data would be made available (after an embargo time) it could be evaluated and published by other scientists and therefore contribute to the entire scientific community.

Even though, open science is on the rise many researchers have not been trained in how to follow the F.A.I.R.+T. principles, and how to make their science and data openly available.
We believe it is paramount to confront scientists at an early career stage with the concepts of open science, and therefore, it was an utmost concern for us to include an entire session on open science in the first ``Czech-Bavarian mini-school on large scale facilities \emph{and open data}'' \cite{mini-school}.
Here, the participants received an introduction to open science \cite{foster}, the F.A.I.R.+T. principle \cite{FAIR} with the complementation of trustworthyness, open publishing \cite{arXiv} and the figshare platform \cite{figshare}, followed by a hands on session.
During the hands on session, openly available data was extracted, evaluated and analysed within approximately one hour, resulting in the data, graphs, and code \cite{data-evaluation} shown below.
We applied the open science principles to triple axis data recorded at IN3 \cite{data} made available by the Institute Laue-Langevin (ILL) in Grenoble.

\section{Experimental details}

The data were recorded using the IN3 triple - axis spectrometer \cite{IN3} at the ILL in 2017.
We are not aware of the details of the experiments, since there was no experimental report or submitted proposal stored together with the data.
The sample measured was a silicon crystal, which is apparent from the sample name chosen in the database, and could be confirmed by the lattice constant of the sample which is 5.431\,\AA \cite{Hom1975}. The sample was oriented such that its reciprocal plane (11$\bar{2}$) lies within the scattering plane of the instrument.

IN3 has two different monochromators: a PG002 and a Cu monochromator.
Considering the d-value of \SI{3.355}{\mbox{\AA}} used in the experiment, which can be extracted from the meta-data, it is clear that the PG002 monochromator was used.
In the same manner it was determined that the PG002 analyser was used. 
Furthermore, the outgoing wave-number $k_f$ was fixed to \SI{2.66325}{\mbox{\AA}^{-1}}, which suggests the use of a PG filter. 
The corresponding wavelength is $\lambda_f=\SI{2.359}{\mbox{\AA}}$.

The sample was cooled down to $T = \SI{1.6}{K}$, where all measurements were recorded. 
This was probably done in the ``orange'' cryostat.

Please note that all numbers used here were extracted from the meta-data ONLY, and we have no way of confirming these data at the moment.

\section{Data analysis and results}

The raw data are analysed utilising Python Jupyter notebooks \cite{jupyter} using the ufit package \cite{ufit}.
%The notebook \cite{data-evaluation} runs within a Docker image \cite{Docker}, that contains all the required Python packages.
%This approach will allow to run the data analysis \cite{data-docker} directly from tools like Binder \cite{binder}.
%Furthermore, it ensures that the data analysis can still be run after a specific package has been upgraded.

ILL, where the data was acquired, is not completely following the F.A.I.R.+T. principles, specifically accessibility. 
In order to access the data one needs to provide ILL credentials. 
Our script is able to download the data from the ILL web page, if the correct ILL credentials are supplied. 
From there the entire raw data folder is downloaded to the /rawdata folder. 
In case the ILL will change their access policy the automatic data download will not work anymore. 
Therefore the raw data directory is published together with the data evaluation scripts \cite{data-evaluation}.

The raw data directory contains 30 data files from number 102942 till 102971. 
The first 27 files are sample alignment, including sample rotation scans, the adjustment of the goniometer and lattice parameters. 
This procedure was repeated several times, possibly as part of a students practice.
Only three ``real'' measurements were performed after the sample alignments (file numbers 10269-12071).
The constant Q scans along the $\Lambda$\,-\,line each contain one excitation, at 10.77\,meV, 13.10\,meV and 14.70\,meV, respectively. 

\begin{figure}
    \includegraphics[width=1.0\linewidth]{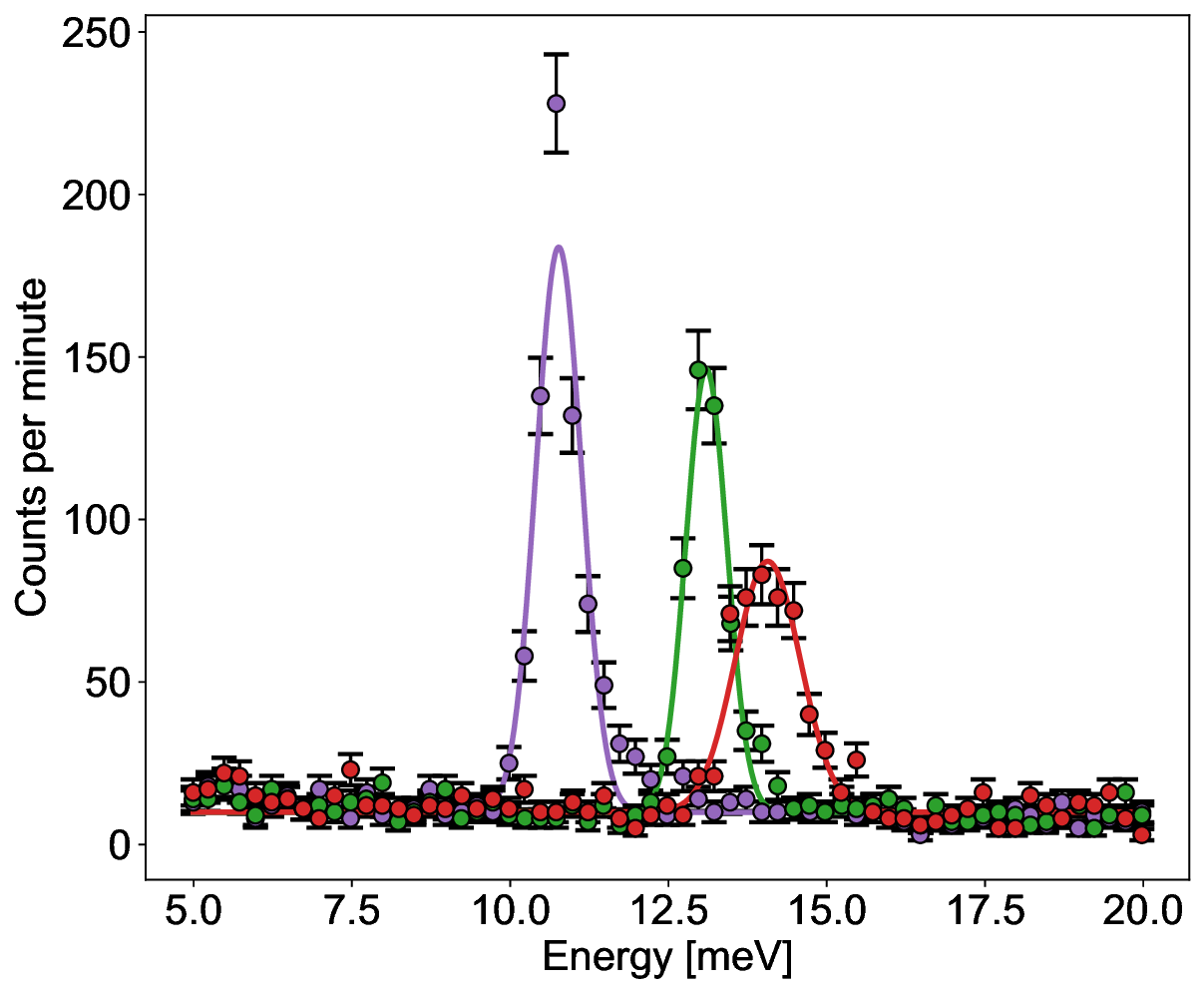}
    \caption{\label{fig1} Constant Q scan along the $\Lambda$\,-\,line, corresponding to the [111] direction.}
\end{figure}

The measured raw data are plotted in the Fig. \ref{fig1}.
Peaks are fitted with a simple Gaussian and a constant background. A detailed analysis of the instrument resolution is beyond the scope of this paper.

\begin{figure}
    \includegraphics[width=1.0\linewidth]{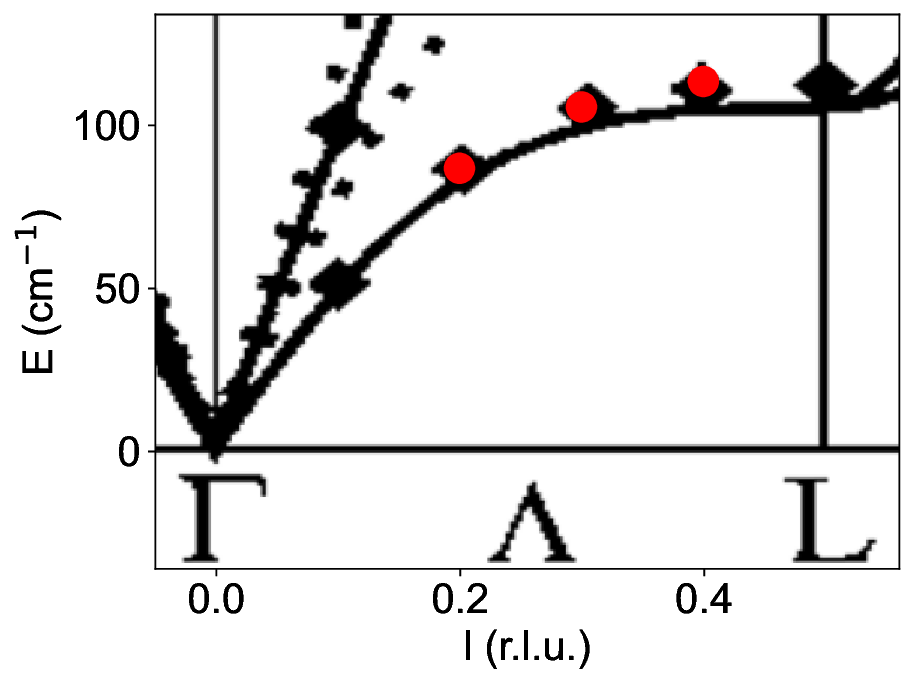}
    \caption{Comparison of the data point we evaluated with the data previously published in \cite{Aouissi} }
    \label{fig2}
\end{figure}

The extracted peak positions are at 10.77\,meV, 13.10\,meV and 14.70\,meV.
These values are plotted as red dots in Fig. \ref{fig2} together with the dispersion relation published in \cite{Aouissi}.

A description on how the data evaluation is run is included in the \emph{readme} file \cite{data-evaluation}.

\section{Conclusions}

We have used freely available resources to extract, evaluate and analyse openly available data, proving that it is possible to follow the FAIR principles with a minimum amount of effort. 
The discrepancy between our results and the previously published data \cite{Aouissi} is due to the fact that we did not consider resolution effects here, since this would go beyond the scope of this exercise.
It needs to be stated that this data analysis would have been much easier, if together with the data, some more information regarding the experiment would have been published, e.g. an experimental report or a lab book.
We would therefore encourage users to add some further description to their openly available data.

\begin{acknowledgments}
We acknowledge all the students who measured neutron data \cite{data} during the Hercules practical course and especially their local contact Paul Steffens. 
We also acknowledge the BTHA agency for providing us with financial support for the Czech Bavarian Mini-school on large-scale facilities and open data under the grant BTHA-SW-2022-4. We acknowledge the Institute Laue-Langevin for opening the data.
\end{acknowledgments}

\bibliography{bibliography}

\begin{thebibliography}{20}%
\makeatletter
\providecommand \@ifxundefined [1]{%
 \@ifx{#1\undefined}
}%
\providecommand \@ifnum [1]{%
 \ifnum #1\expandafter \@firstoftwo
 \else \expandafter \@secondoftwo
 \fi
}%
\providecommand \@ifx [1]{%
 \ifx #1\expandafter \@firstoftwo
 \else \expandafter \@secondoftwo
 \fi
}%
\providecommand \natexlab [1]{#1}%
\providecommand \enquote  [1]{``#1''}%
\providecommand \bibnamefont  [1]{#1}%
\providecommand \bibfnamefont [1]{#1}%
\providecommand \citenamefont [1]{#1}%
\providecommand \href@noop [0]{\@secondoftwo}%
\providecommand \href [0]{\begingroup \@sanitize@url \@href}%
\providecommand \@href[1]{\@@startlink{#1}\@@href}%
\providecommand \@@href[1]{\endgroup#1\@@endlink}%
\providecommand \@sanitize@url [0]{\catcode `\\12\catcode `\$12\catcode
  `\&12\catcode `\#12\catcode `\^12\catcode `\_12\catcode `\%12\relax}%
\providecommand \@@startlink[1]{}%
\providecommand \@@endlink[0]{}%
\providecommand \url  [0]{\begingroup\@sanitize@url \@url }%
\providecommand \@url [1]{\endgroup\@href {#1}{\urlprefix }}%
\providecommand \urlprefix  [0]{URL }%
\providecommand \Eprint [0]{\href }%
\providecommand \doibase [0]{http://dx.doi.org/}%
\providecommand \selectlanguage [0]{\@gobble}%
\providecommand \bibinfo  [0]{\@secondoftwo}%
\providecommand \bibfield  [0]{\@secondoftwo}%
\providecommand \translation [1]{[#1]}%
\providecommand \BibitemOpen [0]{}%
\providecommand \bibitemStop [0]{}%
\providecommand \bibitemNoStop [0]{.\EOS\space}%
\providecommand \EOS [0]{\spacefactor3000\relax}%
\providecommand \BibitemShut  [1]{\csname bibitem#1\endcsname}%
\let\auto@bib@innerbib\@empty
%</preamble>
\bibitem [{\citenamefont {Beddrich}\ \emph {et~al.}(2020)\citenamefont
  {Beddrich}, \citenamefont {Book}, \citenamefont {Brems}, \citenamefont
  {Čermák}, \citenamefont {Dembski-Villalta}, \citenamefont {Flacke},
  \citenamefont {Gabold}, \citenamefont {Gerina}, \citenamefont {Jochum},
  \citenamefont {Kancko}, \citenamefont {Kohúteková}, \citenamefont
  {Košutová}, \citenamefont {Král}, \citenamefont {Murmiliuk}, \citenamefont
  {Nowak}, \citenamefont {Pylypets}, \citenamefont {Staško}, \citenamefont
  {Tang}, \citenamefont {Vančík},\ and\ \citenamefont
  {Vogl}}]{minischool2020}%
  \BibitemOpen
  \bibfield  {author} {\bibinfo {author} {\bibfnamefont {L.}~\bibnamefont
  {Beddrich}}, \bibinfo {author} {\bibfnamefont {A.}~\bibnamefont {Book}},
  \bibinfo {author} {\bibfnamefont {X.~S.}\ \bibnamefont {Brems}}, \bibinfo
  {author} {\bibfnamefont {P.}~\bibnamefont {Čermák}}, \bibinfo {author}
  {\bibfnamefont {M.}~\bibnamefont {Dembski-Villalta}}, \bibinfo {author}
  {\bibfnamefont {L.}~\bibnamefont {Flacke}}, \bibinfo {author} {\bibfnamefont
  {H.}~\bibnamefont {Gabold}}, \bibinfo {author} {\bibfnamefont
  {M.}~\bibnamefont {Gerina}}, \bibinfo {author} {\bibfnamefont {J.~K.}\
  \bibnamefont {Jochum}}, \bibinfo {author} {\bibfnamefont {A.}~\bibnamefont
  {Kancko}}, \bibinfo {author} {\bibfnamefont {S.}~\bibnamefont
  {Kohúteková}}, \bibinfo {author} {\bibfnamefont {T.}~\bibnamefont
  {Košutová}}, \bibinfo {author} {\bibfnamefont {P.}~\bibnamefont {Král}},
  \bibinfo {author} {\bibfnamefont {A.}~\bibnamefont {Murmiliuk}}, \bibinfo
  {author} {\bibfnamefont {L.}~\bibnamefont {Nowak}}, \bibinfo {author}
  {\bibfnamefont {A.}~\bibnamefont {Pylypets}}, \bibinfo {author}
  {\bibfnamefont {D.}~\bibnamefont {Staško}}, \bibinfo {author} {\bibfnamefont
  {R.}~\bibnamefont {Tang}}, \bibinfo {author} {\bibfnamefont {M.}~\bibnamefont
  {Vančík}}, \ and\ \bibinfo {author} {\bibfnamefont {L.}~\bibnamefont
  {Vogl}},\ }\href {https://arxiv.org/abs/2010.12086v2} {\enquote {\bibinfo
  {title} {An exercise in open data: Triple axis data on si single crystal
  (v2)},}\ } (\bibinfo {year} {2020})\BibitemShut {NoStop}%
\bibitem [{fos()}]{foster}%
  \BibitemOpen
  \href@noop {} {\enquote {\bibinfo {title} {Foster open data},}\ }\bibinfo
  {howpublished} {\url{https://www.fosteropenscience.eu/}},\ \bibinfo {note}
  {accessed: 2020-10-22}\BibitemShut {NoStop}%
\bibitem [{\citenamefont {Aroyo}\ \emph {et~al.}(2006)\citenamefont {Aroyo},
  \citenamefont {Kirov}, \citenamefont {Capillas}, \citenamefont {Perez-Mato},\
  and\ \citenamefont {H.}}]{Bilbao}%
  \BibitemOpen
  \bibfield  {author} {\bibinfo {author} {\bibfnamefont {M.~I.}\ \bibnamefont
  {Aroyo}}, \bibinfo {author} {\bibfnamefont {A.}~\bibnamefont {Kirov}},
  \bibinfo {author} {\bibfnamefont {C.}~\bibnamefont {Capillas}}, \bibinfo
  {author} {\bibfnamefont {J.~M.}\ \bibnamefont {Perez-Mato}}, \ and\ \bibinfo
  {author} {\bibfnamefont {W.}~\bibnamefont {H.}},\ }\href@noop {} {\bibfield
  {journal} {\bibinfo  {journal} {Acta Crystallographica}\ }\textbf {\bibinfo
  {volume} {A62}},\ \bibinfo {pages} {115} (\bibinfo {year}
  {2006})}\BibitemShut {NoStop}%
\bibitem [{\citenamefont {NIST}()}]{NIST}%
  \BibitemOpen
  \bibfield  {author} {\bibinfo {author} {\bibnamefont {NIST}},\ }\href@noop {}
  {\enquote {\bibinfo {title} {Nist inorganic crystal structure database, nist
  standard reference database number 3, national institute of standards and
  technology},}\ }\bibinfo {howpublished}
  {\url{https://data.nist.gov/od/id/mds2-2147}}\BibitemShut {NoStop}%
\bibitem [{\citenamefont {Molloy}(2011)}]{plos}%
  \BibitemOpen
  \bibfield  {author} {\bibinfo {author} {\bibfnamefont {J.~C.}\ \bibnamefont
  {Molloy}},\ }\href {\doibase 10.1371/journal.pbio.1001195} {\bibfield
  {journal} {\bibinfo  {journal} {PLOS Biology}\ }\textbf {\bibinfo {volume}
  {9}},\ \bibinfo {pages} {1} (\bibinfo {year} {2011})}\BibitemShut {NoStop}%
\bibitem [{\citenamefont {Fischer}\ and\ \citenamefont
  {Zigmond}(2010)}]{sharing}%
  \BibitemOpen
  \bibfield  {author} {\bibinfo {author} {\bibfnamefont {B.~A.}\ \bibnamefont
  {Fischer}}\ and\ \bibinfo {author} {\bibfnamefont {M.~J.}\ \bibnamefont
  {Zigmond}},\ }\href@noop {} {\bibfield  {journal} {\bibinfo  {journal}
  {Science and engineering ethics}\ }\textbf {\bibinfo {volume} {16}},\
  \bibinfo {pages} {783} (\bibinfo {year} {2010})}\BibitemShut {NoStop}%
\bibitem [{pan()}]{panosc}%
  \BibitemOpen
  \href@noop {} {\enquote {\bibinfo {title} {The photon and neutron open
  science cloud {(PaNOSC)}},}\ }\bibinfo {howpublished}
  {\url{https://www.panosc.eu/}},\ \bibinfo {note} {accessed:
  2020-10-22}\BibitemShut {NoStop}%
\bibitem [{nfd()}]{nfdi}%
  \BibitemOpen
  \href@noop {} {\enquote {\bibinfo {title} {National research data
  infrastructure {(NFDI)}},}\ }\bibinfo {howpublished}
  {\url{https://www.dfg.de/en/research_funding/programmes/nfdi/index.html}},\
  \bibinfo {note} {accessed: 2020-10-22}\BibitemShut {NoStop}%
\bibitem [{exp()}]{expands}%
  \BibitemOpen
  \href@noop {} {\enquote {\bibinfo {title} {European open science cloud
  {(EOSC)} photon and neutron data service {(ExPaNDS)}},}\ }\bibinfo
  {howpublished} {\url{https://expands.eu/}},\ \bibinfo {note} {accessed:
  2020-10-22}\BibitemShut {NoStop}%
\bibitem [{min()}]{mini-school}%
  \BibitemOpen
  \href@noop {} {\enquote {\bibinfo {title} {Czech-bavarian mini-school on
  large scale facilities and open data},}\ }\bibinfo {howpublished}
  {\url{https://mini-school.eu/}},\ \bibinfo {note} {accessed:
  2020-10-22}\BibitemShut {NoStop}%
\bibitem [{FAI()}]{FAIR}%
  \BibitemOpen
  \href@noop {} {\enquote {\bibinfo {title} {{FAIR} principles},}\ }\bibinfo
  {howpublished} {\url{https://www.go-fair.org/fair-principles/}},\ \bibinfo
  {note} {accessed: 2020-10-22}\BibitemShut {NoStop}%
\bibitem [{arX()}]{arXiv}%
  \BibitemOpen
  \href@noop {} {\enquote {\bibinfo {title} {ar{X}iv.org},}\ }\bibinfo
  {howpublished} {\url{https://arxiv.org/}},\ \bibinfo {note} {accessed:
  2020-10-22}\BibitemShut {NoStop}%
\bibitem [{fig()}]{figshare}%
  \BibitemOpen
  \href@noop {} {\enquote {\bibinfo {title} {figshare},}\ }\bibinfo
  {howpublished} {\url{https://figshare.com/}},\ \bibinfo {note} {accessed:
  2020-10-22}\BibitemShut {NoStop}%
\bibitem [{\citenamefont {Baloh}\ \emph {et~al.}(2022)\citenamefont {Baloh},
  \citenamefont {Bauer}, \citenamefont {Bendov\'a}, \citenamefont
  {\v{C}erm\'ak}, \citenamefont {Fellner}, \citenamefont {Ghanathe},
  \citenamefont {\v{S}tefan Hricov}, \citenamefont {Jochum}, \citenamefont
  {Kotvytsk\'a}, \citenamefont {Kumar}, \citenamefont {Labh}, \citenamefont
  {Machovec}, \citenamefont {Pauw}, \citenamefont {Ramszov\'a}, \citenamefont
  {Walz}, \citenamefont {Wild},\ and\ \citenamefont
  {Alvarez}}]{data-evaluation}%
  \BibitemOpen
  \bibfield  {author} {\bibinfo {author} {\bibfnamefont {P.}~\bibnamefont
  {Baloh}}, \bibinfo {author} {\bibfnamefont {L.}~\bibnamefont {Bauer}},
  \bibinfo {author} {\bibfnamefont {A.}~\bibnamefont {Bendov\'a}}, \bibinfo
  {author} {\bibfnamefont {P.}~\bibnamefont {\v{C}erm\'ak}}, \bibinfo {author}
  {\bibfnamefont {K.}~\bibnamefont {Fellner}}, \bibinfo {author} {\bibfnamefont
  {M.}~\bibnamefont {Ghanathe}}, \bibinfo {author} {\bibnamefont {\v{S}tefan
  Hricov}}, \bibinfo {author} {\bibfnamefont {J.~K.}\ \bibnamefont {Jochum}},
  \bibinfo {author} {\bibfnamefont {L.}~\bibnamefont {Kotvytsk\'a}}, \bibinfo
  {author} {\bibfnamefont {S.}~\bibnamefont {Kumar}}, \bibinfo {author}
  {\bibfnamefont {A.}~\bibnamefont {Labh}}, \bibinfo {author} {\bibfnamefont
  {P.}~\bibnamefont {Machovec}}, \bibinfo {author} {\bibfnamefont {B.~R.}\
  \bibnamefont {Pauw}}, \bibinfo {author} {\bibfnamefont {K.}~\bibnamefont
  {Ramszov\'a}}, \bibinfo {author} {\bibfnamefont {E.}~\bibnamefont {Walz}},
  \bibinfo {author} {\bibfnamefont {P.}~\bibnamefont {Wild}}, \ and\ \bibinfo
  {author} {\bibfnamefont {O.~E.~H.}\ \bibnamefont {Alvarez}},\ }\href@noop {}
  {\enquote {\bibinfo {title} {An exercise in open data: Triple axis data on si
  single crystal},}\ }\bibinfo {howpublished}
  {\href{https://doi.org/10.6084/m9.figshare.21407445}{doi:10.6084/m9.figshare.21407445}}
  (\bibinfo {year} {2022})\BibitemShut {NoStop}%
\bibitem [{\citenamefont {Paul}\ \emph {et~al.}(2014)\citenamefont {Paul},
  \citenamefont {Greta}, \citenamefont {Yue}, \citenamefont {Christopher},
  \citenamefont {David}, \citenamefont {Marianna}, \citenamefont {Matthias},
  \citenamefont {Juho}, \citenamefont {Atefeh}, \citenamefont {Emilie},
  \citenamefont {Carolina}, \citenamefont {Hamed}, \citenamefont {Nors},
  \citenamefont {Ramu}, \citenamefont {Umbertoluca}, \citenamefont {Matteo},
  \citenamefont {Sarah}, \citenamefont {Markus}, \citenamefont {Janis},
  \citenamefont {David},\ and\ \citenamefont {Mohamed}}]{data}%
  \BibitemOpen
  \bibfield  {author} {\bibinfo {author} {\bibfnamefont {S.}~\bibnamefont
  {Paul}}, \bibinfo {author} {\bibfnamefont {D.}~\bibnamefont {Greta}},
  \bibinfo {author} {\bibfnamefont {D.}~\bibnamefont {Yue}}, \bibinfo {author}
  {\bibfnamefont {D.}~\bibnamefont {Christopher}}, \bibinfo {author}
  {\bibfnamefont {D.}~\bibnamefont {David}}, \bibinfo {author} {\bibfnamefont
  {G.}~\bibnamefont {Marianna}}, \bibinfo {author} {\bibfnamefont
  {H.}~\bibnamefont {Matthias}}, \bibinfo {author} {\bibfnamefont
  {I.}~\bibnamefont {Juho}}, \bibinfo {author} {\bibfnamefont {J.}~\bibnamefont
  {Atefeh}}, \bibinfo {author} {\bibfnamefont {L.}~\bibnamefont {Emilie}},
  \bibinfo {author} {\bibfnamefont {L.~S.~A.}\ \bibnamefont {Carolina}},
  \bibinfo {author} {\bibfnamefont {P.}~\bibnamefont {Hamed}}, \bibinfo
  {author} {\bibfnamefont {P.~M.}\ \bibnamefont {Nors}}, \bibinfo {author}
  {\bibfnamefont {P.}~\bibnamefont {Ramu}}, \bibinfo {author} {\bibfnamefont
  {R.}~\bibnamefont {Umbertoluca}}, \bibinfo {author} {\bibfnamefont
  {R.}~\bibnamefont {Matteo}}, \bibinfo {author} {\bibfnamefont
  {S.}~\bibnamefont {Sarah}}, \bibinfo {author} {\bibfnamefont
  {S.}~\bibnamefont {Markus}}, \bibinfo {author} {\bibfnamefont
  {T.}~\bibnamefont {Janis}}, \bibinfo {author} {\bibfnamefont
  {V.}~\bibnamefont {David}}, \ and\ \bibinfo {author} {\bibfnamefont
  {Z.}~\bibnamefont {Mohamed}},\ }\href {\doibase 10.5291/ILL-DATA.TEST-2385}
  {\enquote {\bibinfo {title} {Hsc17 hercules practical course},}\ } (\bibinfo
  {year} {2014})\BibitemShut {NoStop}%
\bibitem [{IN3()}]{IN3}%
  \BibitemOpen
  \href@noop {} {\enquote {\bibinfo {title} {{IN3}},}\ }\bibinfo {howpublished}
  {\url{https://www.ill.eu/users/instruments/instruments-list/in3/description/instrument-layout/}},\
  \bibinfo {note} {accessed: 2020-10-22}\BibitemShut {NoStop}%
\bibitem [{\citenamefont {Hom}\ \emph {et~al.}(1975)\citenamefont {Hom},
  \citenamefont {Kiszenik},\ and\ \citenamefont {Post}}]{Hom1975}%
  \BibitemOpen
  \bibfield  {author} {\bibinfo {author} {\bibfnamefont {T.}~\bibnamefont
  {Hom}}, \bibinfo {author} {\bibfnamefont {W.}~\bibnamefont {Kiszenik}}, \
  and\ \bibinfo {author} {\bibfnamefont {B.}~\bibnamefont {Post}},\ }\href
  {\doibase 10.1107/S0021889875010965} {\bibfield  {journal} {\bibinfo
  {journal} {Journal of Applied Crystallography}\ }\textbf {\bibinfo {volume}
  {8}},\ \bibinfo {pages} {457} (\bibinfo {year} {1975})}\BibitemShut {NoStop}%
\bibitem [{\citenamefont {Kluyver}\ \emph {et~al.}(2016)\citenamefont
  {Kluyver}, \citenamefont {Ragan-Kelley}, \citenamefont {P{\'e}rez},
  \citenamefont {Granger}, \citenamefont {Bussonnier}, \citenamefont
  {Frederic}, \citenamefont {Kelley}, \citenamefont {Hamrick}, \citenamefont
  {Grout}, \citenamefont {Corlay}, \citenamefont {Ivanov}, \citenamefont
  {Avila}, \citenamefont {Abdalla},\ and\ \citenamefont {Willing}}]{jupyter}%
  \BibitemOpen
  \bibfield  {author} {\bibinfo {author} {\bibfnamefont {T.}~\bibnamefont
  {Kluyver}}, \bibinfo {author} {\bibfnamefont {B.}~\bibnamefont
  {Ragan-Kelley}}, \bibinfo {author} {\bibfnamefont {F.}~\bibnamefont
  {P{\'e}rez}}, \bibinfo {author} {\bibfnamefont {B.}~\bibnamefont {Granger}},
  \bibinfo {author} {\bibfnamefont {M.}~\bibnamefont {Bussonnier}}, \bibinfo
  {author} {\bibfnamefont {J.}~\bibnamefont {Frederic}}, \bibinfo {author}
  {\bibfnamefont {K.}~\bibnamefont {Kelley}}, \bibinfo {author} {\bibfnamefont
  {J.}~\bibnamefont {Hamrick}}, \bibinfo {author} {\bibfnamefont
  {J.}~\bibnamefont {Grout}}, \bibinfo {author} {\bibfnamefont
  {S.}~\bibnamefont {Corlay}}, \bibinfo {author} {\bibfnamefont
  {P.}~\bibnamefont {Ivanov}}, \bibinfo {author} {\bibfnamefont
  {D.}~\bibnamefont {Avila}}, \bibinfo {author} {\bibfnamefont
  {S.}~\bibnamefont {Abdalla}}, \ and\ \bibinfo {author} {\bibfnamefont
  {C.}~\bibnamefont {Willing}},\ }in\ \href@noop {} {\emph {\bibinfo
  {booktitle} {Positioning and Power in Academic Publishing: Players, Agents
  and Agendas}}},\ \bibinfo {editor} {edited by\ \bibinfo {editor}
  {\bibfnamefont {F.}~\bibnamefont {Loizides}}\ and\ \bibinfo {editor}
  {\bibfnamefont {B.}~\bibnamefont {Schmidt}}}\ (\bibinfo {organization} {IOS
  Press},\ \bibinfo {year} {2016})\ pp.\ \bibinfo {pages} {87 --
  90}\BibitemShut {NoStop}%
\bibitem [{ufi()}]{ufit}%
  \BibitemOpen
  \href@noop {} {\enquote {\bibinfo {title} {Ufit},}\ }\bibinfo {howpublished}
  {\url{https://wiki.mlz-garching.de/ufit:index}},\ \bibinfo {note} {accessed:
  2020-10-22}\BibitemShut {NoStop}%
\bibitem [{\citenamefont {Aouissi}\ \emph {et~al.}(2006)\citenamefont
  {Aouissi}, \citenamefont {Hamdi}, \citenamefont {Meskini},\ and\
  \citenamefont {Qteish}}]{Aouissi}%
  \BibitemOpen
  \bibfield  {author} {\bibinfo {author} {\bibfnamefont {M.}~\bibnamefont
  {Aouissi}}, \bibinfo {author} {\bibfnamefont {I.}~\bibnamefont {Hamdi}},
  \bibinfo {author} {\bibfnamefont {N.}~\bibnamefont {Meskini}}, \ and\
  \bibinfo {author} {\bibfnamefont {A.}~\bibnamefont {Qteish}},\ }\href
  {\doibase 10.1103/PhysRevB.74.054302} {\bibfield  {journal} {\bibinfo
  {journal} {Phys. Rev. B}\ }\textbf {\bibinfo {volume} {74}},\ \bibinfo
  {pages} {054302} (\bibinfo {year} {2006})}\BibitemShut {NoStop}%
\end{thebibliography}%

\end{document}